\documentclass[9pt, twocolumn,twoside]{pnas-new}

\templatetype{pnasresearcharticle} 

\title{Fast-freezing kinetics inside a droplet impacting on a cold surface}

\author[a,1]{Pallav Kant}
\author[a,b]{Robin B. J. Koldeweij}
\author[a]{Kirsten Harth}
\author[a,c]{Michiel A. J. van Limbeek}
\author[a,c,1]{Detlef Lohse}

\affil[a]{Physics of Fluids Group, Max Planck Center Twente for Complex Fluid Dynamics and J. M. Burgers Centre for Fluid Mechanics, MESA+ Institute for Nanotechnology, University of Twente, 7500 AE Enschede, The Netherlands}
\affil[b]{Nano-Instrumentation, TNO, 5612 AP Eindhoven, The Netherlands}
\affil[c]{Max Planck Institute for Dynamics and Self-Organization, 37077 G$\ddot{o}$ttingen, Germany}

\leadauthor{Kant} 

\significancestatement{From critically affecting the performance of an aircraft to the droplet-based additive manufacturing, the solidification of impacting droplets influences a wide range of industrial applications; thus a deeper fundamental understanding of the solidification of an impacting droplet is necessary.
In this work, we reveal and rationalize a peculiar freezing morphology originating from the complex interplay between the droplet scale hydrodynamics and phase-transition effects at sufficiently high substrate undercooling.
The direct visualization of different freezing morphologies only became possible as we adopt a new optical technique (TIR) in the context of freezing.
This technique can be employed in more complex situations, such as solidification of an impacting droplet on liquid infused or patterned surfaces, which potentially can influence many industrial processes.
}

\authorcontributions{Author contributions: R. K., M. v L. and D. L. initiated the research; P. K., R. K. and K. H. performed experiments. P. K. analysed the data and developed the theoretical framework. P. K. and D. L. wrote the manuscript with contributions from R. K. and K. H.}
\authordeclaration{No conflict of interests to be declared.}
\correspondingauthor{\textsuperscript{1}To whom correspondence should be addressed. E-mail: p.kant@utwente.nl and d.lohse@utwente.nl}


\begin{abstract}
Freezing or solidification of impacting droplets is omnipresent in nature and technology, be it a rain droplet falling on a supercooled surface, be it in inkjet printing where often molten wax is used, be it in added manufacturing or in metal production processes or in extreme ultraviolet lithography (EUV) for the chip production where molten tin is used to generate the EUV radiation. 
For many of these industrial applications, a detailed understanding of the solidification process is essential.
Here, by adopting a totally new optical technique in the context of freezing, namely TIR (Total-Internal-Reflection), we elucidate the freezing kinetics during the solidification of a droplet while it impacts on an undercooled surface.
We show for the first time that at sufficiently high undercooling a peculiar freezing morphology exists that involves sequential advection of frozen fronts from the centre of the droplet to its boundaries.
This phenomenon is examined by combining elements of classical nucleation theory to the large scale hydrodynamics on the droplet scale, bringing together two subfields which traditionally have been quite separated.
Furthermore, we report a peculiar self-peeling phenomenon of a frozen splat that is driven by the existence of a transient crystalline state during solidification.
\end{abstract}

\dates{This manuscript was compiled on \today}
\doi{\url{www.pnas.org/cgi/doi/10.1073/pnas.XXXXXXXXXX}}

\begin{document}

\maketitle
\thispagestyle{firststyle}
\ifthenelse{\boolean{shortarticle}}{\ifthenelse{\boolean{singlecolumn}}{\abscontentformatted}{\abscontent}}{}

\dropcap{I}mpact of a droplet on an undercooled solid surface instigates a number of physical processes simultaneously, including drop scale fluid motion, heat transfer between the liquid and the substrate, and the related phase transition.
Whereas a large number of studies have investigated the corresponding interface deformations and the spreading of a droplet after it impinges onto an undercooled surface \cite{madejski1976solidification, Pasandideh-Fard1996, schremb2018normal, bennett1993splat, alizadeh2012dynamics, jin2017experimental, dhiman2005freezing, mishchenko2010design, Tin_JFM_Submitted}, the \emph{kinetics} of phase transition within the impacting droplet has been addressed only in a few \cite{schremb2017transient, wang2019effect, schremb2017ice}.
Moreover, among the studies concerning solidification kinetics, only the regimes where phase transition effects are slower than the fast dynamics of droplet impact have been investigated.
Here, we explore freezing behaviours that arise due to the rapid solidification of an impacting droplet at a sufficiently high substrate undercooling.
Such scenarios are encountered in a number of industrial processes ranging from additive manufacturing \cite{gao1994precise, visser20153d} to thermal plasma spraying of ceramics and metallic materials \cite{lawley1994spray,evans1985osprey,liu2001high}, extreme ultraviolet lithography \cite{mizoguchi2011100w, banine2011physical} etc.

In the present work, we adapt the total-internal-reflection (TIR) technique \cite{kolinski2012skating, shirota2017measuring, shirota2016dynamic} to visualise the phase transition in the vicinity of the liquid-substrate interface after a droplet impacts onto an undercooled transparent surface.
This unique technique allows temporally and spatially resolved insight into the nucleation events and crystal growth occurring next to the cold surface on an evanescent length scale ($\sim$ 100 nm), which is otherwise inaccessible through any
other optical technique.
Moreover, it allows us to monitor the delamination of the frozen-splat from the substrate at later time.
The details of the experimental setup are provided in the Materials and Methods section.

\begin{figure*}
\centering
\includegraphics[clip, trim=0cm 0cm 0cm 0cm, width=1.0\textwidth]{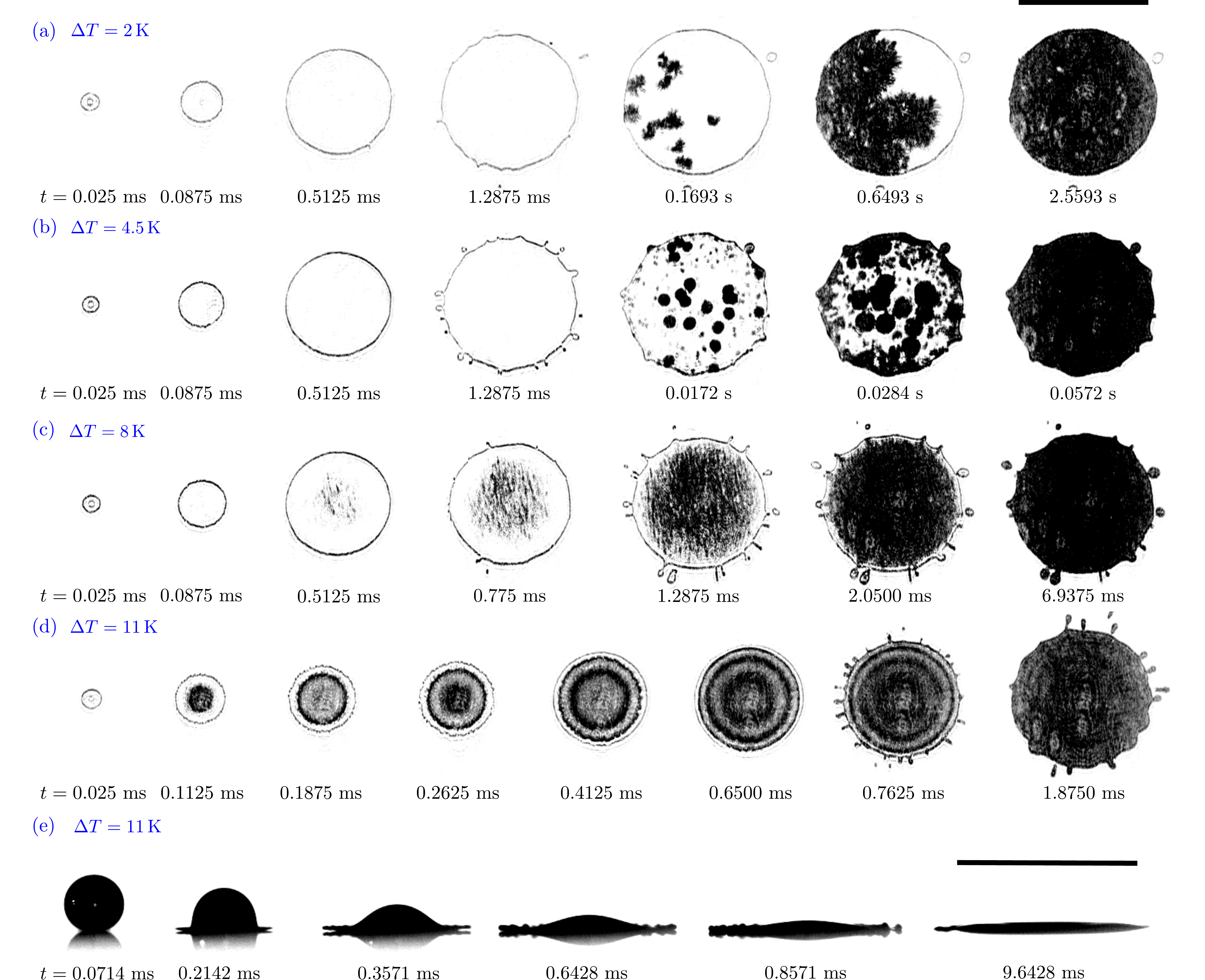}\\
\caption{(a-d) Sequences of experimental snapshots highlighting the effect of substrate undercooling $\Delta T= T_{\mathrm{m}} - T_{\mathrm{s}}$= $\mathrm{(a)}\,2\,\mathrm{K}~\mathrm{(b)}\,4.5\,\mathrm{K}~\mathrm{(c)}\,8\, \mathrm{K}~ \mathrm{(d)}\,11\,\mathrm{K}$ on the freezing morphology of an impacting droplet. In all the cases the velocity and the temperature of the droplet at the time of impact are $U = 2.8$ m/s and $T_\mathrm{d} = 20^\circ$C, respectively. (e) Typical interface deformations associated with the impact of a droplet of size $D = 1.58$ mm on an undercooled surface; recorded in side view at 14000 fps. The scale bars shown on the images correspond to 5 mm. Corresponding movies (S1 -- S4) are available online as supplementary material.}
\label{fig:fig1}
\end{figure*}

\section*{Results and Discussion}
The sequences of experimental snapshots in Fig. \ref{fig:fig1}(a-d) highlight different freezing morphologies of a liquid hexadecane droplet after it impinges on a flat cold surface at various undercooling $\Delta T = T_{\mathrm{m}} - T_{\mathrm{s}}$; $T_{\mathrm{m}} = 18^\circ$C is the melting temperature of the liquid and $T_{\mathrm{s}}$ is the temperature of the substrate.
In all the cases, a droplet of diameter $D = 1.58\,\pm 0.02\,\mathrm{mm}$ at ambient temperature $T_\mathrm{d} \approx 20^\circ$C impacts onto the horizontal surface of an undercooled sapphire prism with a velocity $U \simeq 2.8$ m/s. 
The Weber number $We = \rho D U^2/\sigma$ and Reynolds numbers $Re = \rho D U/ \mu$ associated with this impact are 340 and 1150, respectively; $\rho$ is the density, $\sigma$ is the surface tension and $\mu$ is the dynamic viscosity of the liquid.
Typical deformations of the droplet interface, recorded from side-view, associated with these impact parameters are shown in the sequence Fig. \ref{fig:fig1}e.
Note that, since hexadecane highly wets the sapphire surface, with static equilibrium contact angle $\theta_{\mathrm{e}} \approx 15^\circ$, no receding or rebound motion of droplets was observed after the impact.

Immediately after a droplet makes contact with the flat substrate two concentric rings appear in the TIR images; see the first panel of each sequence in Fig. \ref{fig:fig1}(a-d).
The outer ring corresponds to the contact line of the droplet that continues to move radially outwards.
The inner ring highlights the footprint of the bubble entrapped underneath the droplet interface \cite{chandra1991collision, thoroddsen2005air, josserand2016drop}.
For low undercooling $\Delta T \leq 8$ K, the inner ring disappears from the images for $t > 0.0375$ ms, suggesting that the entrapped bubble detaches from the substrate and rises within the liquid.
This detachment is directly related to the strong wetting of sapphire by hexadecane, as described in Ref. \cite{san2012does}.
For higher undercooling $\Delta T > 8$ K, on the contrary, the entrapped bubble freezes in contact with the substrate and at later times acts as a localised defect that causes delamination of the frozen splat; see supplementary movie.

For droplet impacts on surfaces maintained at low undercooling, the phase transition does not initiate instantly after a droplet touches the cold substrate.
Instead, the crystal nucleation occurs randomly on the wetted area at timescales that are much larger than the typical timescale $t_{\mathrm{I}} \sim D/U$ during which the impact driven dynamics last. 
We believe, this is related to the high energy barrier for the formation of a stable crystal at temperatures close to the melting point\cite{turnbull1952kinetics, kurz1986fundamentals}.
Therefore at low $\Delta T$, facilitated by various inhomogeneities present in the system (surface roughness, thermal fluctuations, impurities in the liquid etc.) predominantly heterogeneous nucleation takes place.
This also corroborates with the observed non-uniformity in the crystal seeding locations at small undercooling.
For example, at $\Delta T = 2 \mathrm{K}$ (Fig. \ref{fig:fig1}b),  after the impact, the droplet spreads out in the form of a pancake of maximum diameter $D_{\mathrm{f}}$ until $t  \sim 3$ ms, whereas the induction time of crystals is much larger $t_{\mathrm{d}} \sim 40$ ms.
With crystal induction time we refer to the time required to form a stable nucleus and its growth to the detectable size \cite{sohnel1988interpretation}. 
In this case, after the nucleation at a few sites ($N \sim 10$), the crystallites grow in the form of needle-like structures (columnar dendrites), which eventually cover the whole wetted area of the substrate at $t \sim 2 \times 10^{3}$ ms.
Note that the coverage of the wetted area by dendrites only indicates the end of the first phase of droplet freezing during which the liquid-substrate interface solidifies \cite{schremb2017ice}.
The bulk freezing of the liquid pancake takes place at much longer timescales $t_{\mathrm{f}} \sim \rho_{\mathrm{f}} L h^2/k_\mathrm{f}\,\Delta T \sim \mathcal{O}(10)$ s; $\rho_\mathrm{f}$ and $k_\mathrm{f}$ are the density and the thermal conductivity of the solidified hexadecane, $L$ is latent heat released per unit mass, $h$ is the thickness of the frozen splat.

We noted that a slight increase in the substrate undercooling significantly influences the overall freezing morphology as well as the timescales associated with it.
At $\Delta T = 4.5$ K (Fig. \ref{fig:fig1}b), the induction time of crystals noticeably reduces to $t_{\mathrm{d}} \sim 5\,\mathrm{ms} > t_{\mathrm{I}}$.
Besides, the nucleation occurs at a far larger number of locations ($N \sim \mathcal{O}(100)$) randomly distributed over the wetted area.
Furthermore, in stark contrast to the spatially erratic growth of columnar dendrites at $\Delta T = 2$ K, here, the crystallites grow uniformly with nearly circular footprints in the plane parallel to the cold surface.
However, a close inspection of the images reveals that the dendritic patterns persist within these uniformly growing crystallites.
This suggests that the crystal grains assume equiaxed dendritic morphology \cite{kurz1986fundamentals} at slightly increased $\Delta T$.

Strikingly, these dendritic morphologies were absent for droplets impacting on surfaces at $\Delta T \geq 8$ K.
For instance, while a droplet spreads on a surface maintained at $\Delta T = 8$ K (Fig. \ref{fig:fig1}c), the nucleation initiates at an enhanced rate in the middle of the evolving wetted area where the liquid temperatures are the lowest (Since the low thermal effusivity $e_{\mathrm{d}} = \sqrt{k \, \rho \, c_\mathrm{p}}$ of hexadecane limits the rate of heat removal from the warm droplet, the temperatures within the droplet are the lowest in the region that remains in contact with the cold substrate for the longest time.).
As a consequence, at $t_{\mathrm{d}} \sim 0.5$ ms $< t_{\mathrm{I}}$, a cloud of micron-sized crystallites forms in the central region.
Subsequently, this cloud grows nearly axi-symmetrically and covers the whole wetted area shortly after the droplet reaches its final resting size at $t \sim 3$ ms.
Note that, at such high undercooling, the steep increase in the nucleation rate is linked to the corresponding decline in the energy barrier for the formation of a stable nucleus at higher $\Delta T$.
This, in turn, reduces the dependence on the extrinsic factors for nucleation and induces prompt nucleation at an enhanced rate within the cooled liquid next to the substrate.
In the following, we will see that the immediate nucleation at even higher $\Delta T$ leads to an unusual freezing morphology and dynamics, which to the best of our knowledge has not been reported earlier.

\begin{figure*}
\centering
\includegraphics[clip, trim=0cm 0cm 0cm 0cm, width=0.8\textwidth]{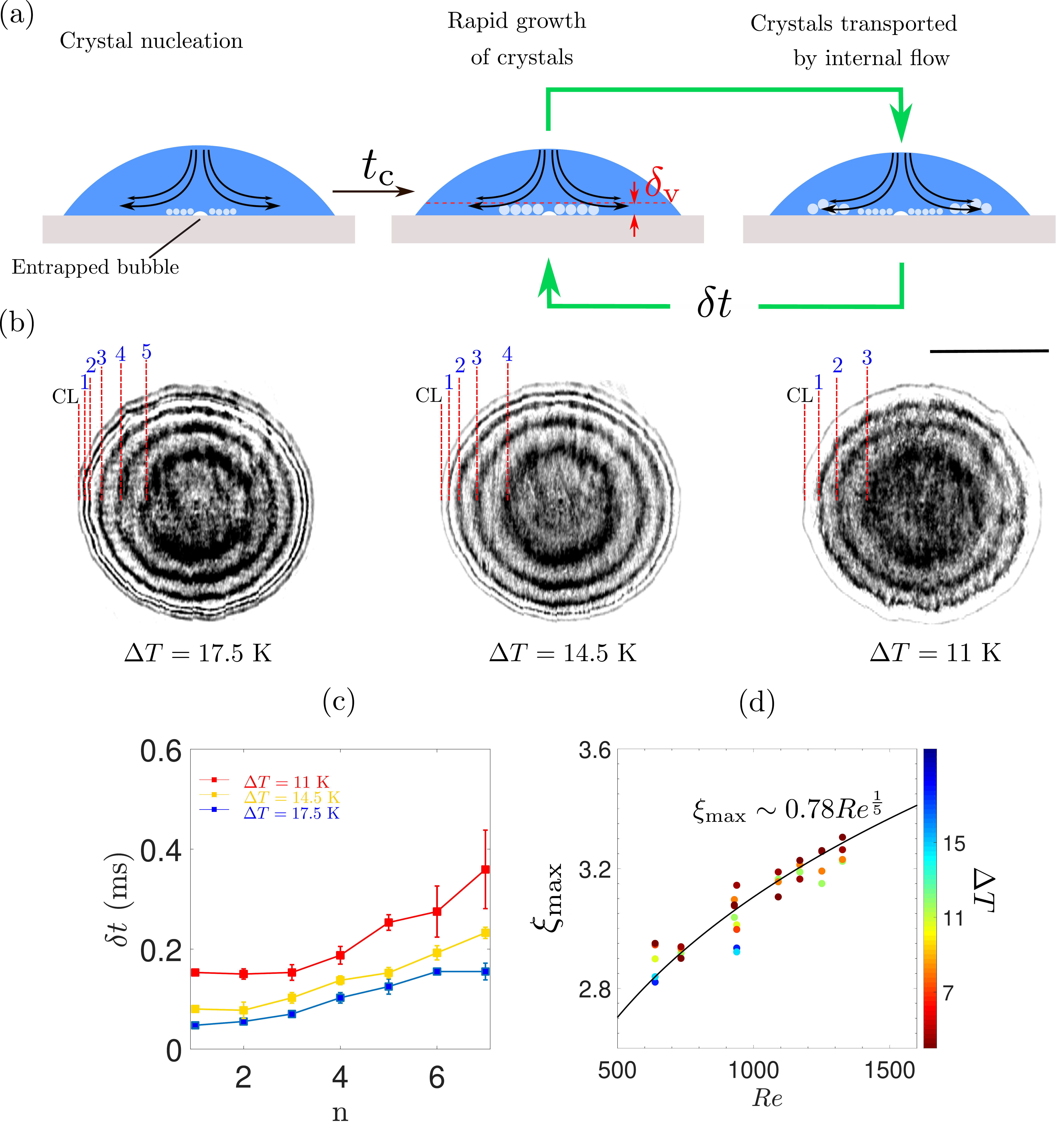}\\
\caption{(a) Schematic representation of the physical mechanism responsible for the sequential advection of frozen fronts during the impact of a droplet on an undercooled surface with $\Delta T \geq 11$ K. (b) Three experimental snapshots recorded at $t = 0.5$ ms after the impacting droplet ($U = 1.7$ m/s) makes contact with the surfaces maintained at $\Delta T$ = 17.5 K, 14.5 K and 11 K. The higher the $\Delta T$, the quicker the advection of frozen fronts. At $\Delta T = 17.5$ K and 14.5 K the advected crystallites accumulate near the  advancing contact line (CL) during early stages of the droplet spreading after impact, however CL always precedes the advected frozen fronts. The horizontal scale bar over the image correspond to 2 mm. Corresponding movie (S5) is available online as supplementary material (c) Variation of time-interval $\delta t$ for successive frozen fronts (n)  (d) Influence of freezing on the maximum spread of the droplet. In our system, the non-dimensional maximum spreading factor $\xi_{\mathrm{max}} = D_{\mathrm{f}}/D$ follows the scaling derived for iso-thermal impact, $\xi_{\mathrm{max}} \sim 0.78 Re^{\frac{1}{5}}$ (solid line).}
\label{fig:fig2}
\end{figure*}

For $\Delta T = 11$ K (Fig. \ref{fig:fig1}d), we observed that shortly after the impact of the drop, a cloud of crystallites appears in the middle of the wetted area at $t_{\mathrm{d}} \sim 0.112$ ms.
The footprint of this cloud grows in size like in the previous case, however, at $t \sim 0.162$ ms, this cloud suddenly evolves into a circular band (frozen front) that moves radially outwards.
Subsequently, while this frozen front moves towards the advancing contact line of the droplet, a second cloud emerges in the middle of the wetted area at $t \sim 0.275$ ms.
This cloud as well evolves into a frozen front that moves out from the central region.
Interestingly, this process repeats a number of times resulting in the sequential advection of several frozen fronts before the droplet-substrate interface completely solidifies at $t \sim 3.625$ ms.
It is important to point out that the sequential advection of the frozen fronts does not occur at a fixed interval of time $\delta t$.
In time this interval increases monotonically.
We measured that the initial two frozen fronts move out from the central region in the intervals of $\delta t \sim 0.1625$ ms, whereas the subsequent frozen fronts appear at $\delta t > 0.1625$ ms; see Fig. \ref{fig:fig2}c.
This striking increase in the time interval for the crystallite front emission is addressed in detail later in the discussion.
Furthermore, we noted that the impact velocity has non-significant influence on $\delta t$. 
This was confirmed by varying the Weber number in between $We =$ 120--450.

The physical mechanism responsible for this unusual freezing morphology is outlined schematically in Fig. \ref{fig:fig2}a.
We hypothesize that while a droplet spreads after the impact onto a highly undercooled surface, a thin layer of liquid next to the cold substrate rapidly cools down.
Consequently, immediate nucleation occurs within this cooled liquid layer as well as on the solid-liquid interface.
The nucleation first initiates near the location of the droplet impact.
Subsequently, the nucleated crystals upon growing to a critical size ($\mathrm{r}_c \sim \delta_v/2$, the thickness of the viscous boundary layer) at $t = t_{\mathrm{c}} \ll t_{\mathrm{I}}$ are advected by the radial flow inside the droplet to form a frozen front.
The whole process repeats each time the crystals nucleated in the middle of the wetted area are swept by the radial flow inside the droplet, leading to the sequential advection of the frozen fronts.
Therefore, the growth rate of crystals within the viscous boundary layer next to the substrate dictates the overall freezing morphology by controlling the timescale at which the frozen fronts are advected.
Since the growth of crystals in the vicinity of the undercooled surface strongly depends on the temperature of the substrate-liquid interface ($T_{\mathrm{sl}}$), it arises as the most crucial parameter of our system.

Qualitatively, the overall freezing morphology does not change for a further increase in $\Delta T$; see Fig. \ref{fig:fig2}b.
However, lowering of the substrate temperature significantly influences the two important timescales associated with the freezing phenomenology, namely the induction time $t_{\mathrm{d}}$ and the periodicity $\delta t$.
For experiments performed at $\Delta T = 14.5$ K and $\Delta T = 17.5$ K, the  induction time drops down to lower values of $t_{\mathrm{d}} = 0.0625$ ms and 0.0375 ms, respectively.
Similarly, the $\delta t$ for the initial frozen fronts in these cases reduces to 0.0875 ms and 0.050 ms, respectively; see Fig. \ref{fig:fig2}c.
The prompt inception of the frozen fronts in these cases leads to the accumulation of crystals near the advancing contact line at  early stages of spreading after impact, $t < 0.5 D/U \sim \mathcal{O}(0.5)$ ms.
However, we noted that the contact line always precedes the advected frozen fronts and the subsequent spreading of the droplet remains unaffected; see supplementary movie.
Accordingly, the non-dimensional maximum spreading diameter of the liquid pancake after drop impact follows the scaling derived for iso-thermal impacts, $\xi_{\mathrm{max}} = D_{\mathrm{f}}/D \sim 0.78 Re^{\frac{1}{5}}$ \cite{roisman2009inertia, wildeman2016spreading}; see Fig. \ref{fig:fig2}d.
This indicates that the kinetic energy loss due to the solidification is too small to affect the extent of droplet spreading.
This is further re-affirmed by the large values of the Prandtl number $Pr = \mu/\rho\,\kappa_{\mathrm{l}} \gg 1$ for liquid hexadecane, which signifies that the viscous effects dominate over the thermal effects during the impact of a droplet onto a cold surface; $\kappa_\mathrm{l}$ is the thermal diffusivity of liquid hexadecane.

More quantitative insight into the sequential advection of frozen fronts can be gained from the theoretical analysis presented below.
Firstly, using classical nucleation theory we estimate the growth rate of crystals ($\mathcal{V}$) in the vicinity of the substrate-liquid interface at temperature $T_{\mathrm{sl}}$.
This yields:
\begin{equation}
\mathcal{V} = \frac{\mathrm{d}r}{\mathrm{d}t} = \frac{D}{\lambda}\,\bigg[1 - \mathrm{exp}\left\{  -{\frac{M\, \Delta G_{f,v}}{\rho N_{\mathrm{A}}\, k_\mathrm{B}\,T_{\mathrm{sl}}}} \right\} \bigg],
\label{eq:eq1}
\end{equation}
where, $M = 226.448\,\mathrm{g/mol}$ is the molar mass of hexadecane, $N_{\mathrm{A}}$ is the Avogadro constant, $k_B$ is the Boltzmann coefficient, $\lambda$ is the mean free path of a liquid molecule, $D$ is the diffusion coefficient, $\Delta G_{f,v} = L\,(T_{\mathrm{m}} - T_{\mathrm{sl}})/T_{\mathrm{m}}$ is the volumetric free energy difference between solid and liquid.
The derivation of the growth rate of a crystal is provided in the supplementary material.
Note that the temperature of the solid-liquid interface $T_{\mathrm{sl}}$ in Eq. (\ref{eq:eq1}) is unknown.
Here, we adopt two separate strategies to estimate this temperature.
These provide appropriate estimates of $T_\mathrm{sl}$ during the early and the later stages of droplet impact.

Since, at early times, only a small fraction of liquid in the middle of the wetted area transforms into crystallites, we ignore the latent heat released during the solidification.
This allows us to approximate the interface temperature $T_\mathrm{sl}$ as the contact temperature between two semi-infinite bodies (droplet and substrate) that are bought into contact \cite{carslaw1959conduction}, as:
\begin{equation}
T_{\mathrm{sl}} = T_{\mathrm{s}} + (T_{\mathrm{d}} - T_{\mathrm{s}})\, \frac{1}{1+e_{\mathrm{s}}/e_{\mathrm{d}}},
\label{eq:eq2}
\end{equation}
$e = \sqrt{\kappa\, \rho \, c_\mathrm{p}}$ is the thermal effusivity, where the subscripts $s$ and $d$ denote the substrate and droplet, respectively. 
In contrast, at later times, due to the continuous nucleation and growth of crystallites a significant amount of latent heat is released near the substrate (Stefan number $St =c_\mathrm{p}\,\Delta T/ L < 1$).
This results in the localised heating of the substrate.
Therefore, to estimate $T_{\mathrm{sl}}$ at later times, we solve the two-phase Stefan problem with phase change using the \textsl{Schwarz}-solution \cite{carslaw1959conduction, loulou1997interface}.
The details of this model are provided in the supplementary material.

\begin{figure}
\centering
\includegraphics[clip, trim=0cm 0cm 0cm 0cm, width=0.35\textwidth]{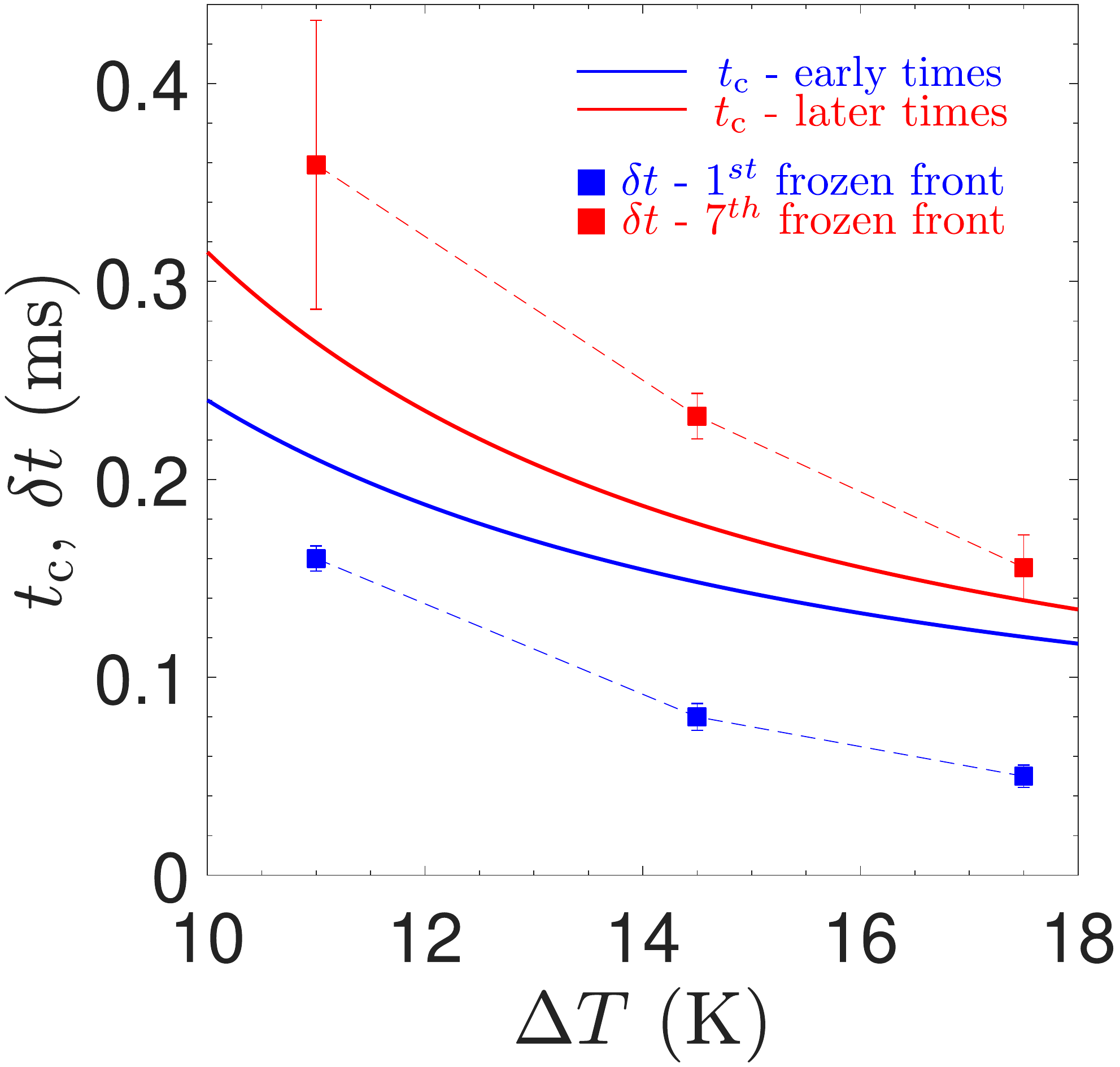}\\
\caption{Comparison between the crossover time $t_\mathrm{c}$ determined from the model and the interval time $\delta t$ between successive frozen fronts measured in the experiments at different $\Delta T$. Blue (red) symbols indicate the experimental $\delta t$ for frozen fronts at early (late) stages of the droplet spreading. Blue (red) solid line highlight the prediction from the model using the substrate-liquid interface temperature estimated without (with) including latent heat released during solidification at early (late) stages of droplet spreading.}
\label{fig:fig3}
\end{figure}

According to Eq. \ref{eq:eq1}, for a fixed value of $T_\mathrm{sl}$, a crystal grows linearly in time ($r_\mathrm{c} \propto t$), whereas, the thickness of the viscous boundary layer inside an impacting droplet grows as $\delta_{\mathrm{v}} \sim 1.88\,\sqrt{\nu t}$ \cite{roisman2009inertia}.
Therefore, a crossover occurs when the crystal size becomes comparable to half the thickness of the viscous boundary layer at
\begin{equation}
t_{\mathrm{c}} \sim 0.88\, \frac{\nu}{\mathcal{V}^2}.
\label{eq:eq2}
\end{equation}
Since we assume that the growing crystals are transported instantly after they reach the critical size ($r_{\mathrm{c}} \sim \delta_{\mathrm{v}}/2$), the crossover time $t_\mathrm{c}$ provides an estimate of the time interval between two successive frozen fronts. 
A comparison between the time interval $t_{\mathrm{c}}$ computed from our model at different $\Delta T$ and the corresponding  experimental measurements of $\delta t$ is shown in Fig. \ref{fig:fig3}.
Despite the sensitivity of the model on various microscopic parameters that are not easily quantified, the model predictions for $\delta t$ are in reasonable agreement with the experimentally measured values.
Our model successfully estimates the low values of $\delta t$ during the early times as well as the delay in the periodicity at later times.
This also confirms that the rise in $\delta t$ at later times is related to the increase in $T_\mathrm{sl}$ due to the release of latent heat during solidification which slows down the growth rate of crystals.
However, note that at later times, a complex interplay between the thickening of the viscous boundary layer and slow growth rate of crystals determines $\delta t$, which is not included in the model.
We must also point out that the deviation between the model predictions and the experimental measurements increases for larger $\Delta T$.
We suspect this difference arises due to the non-equilibrium effects during the impact of a droplet.
In our model we only include the slow diffusion-controlled growth of crystals in a quiescent undercooled liquid, which may not be an accurate description of the conditions during the nucleation at the early stages of the droplet impact.
An accurate estimation would require a detailed modelling of the system which is beyond the scope of this work.

\begin{figure}
\centering
\includegraphics[clip, trim=0cm 0cm 0cm 0cm, width=0.5\textwidth]{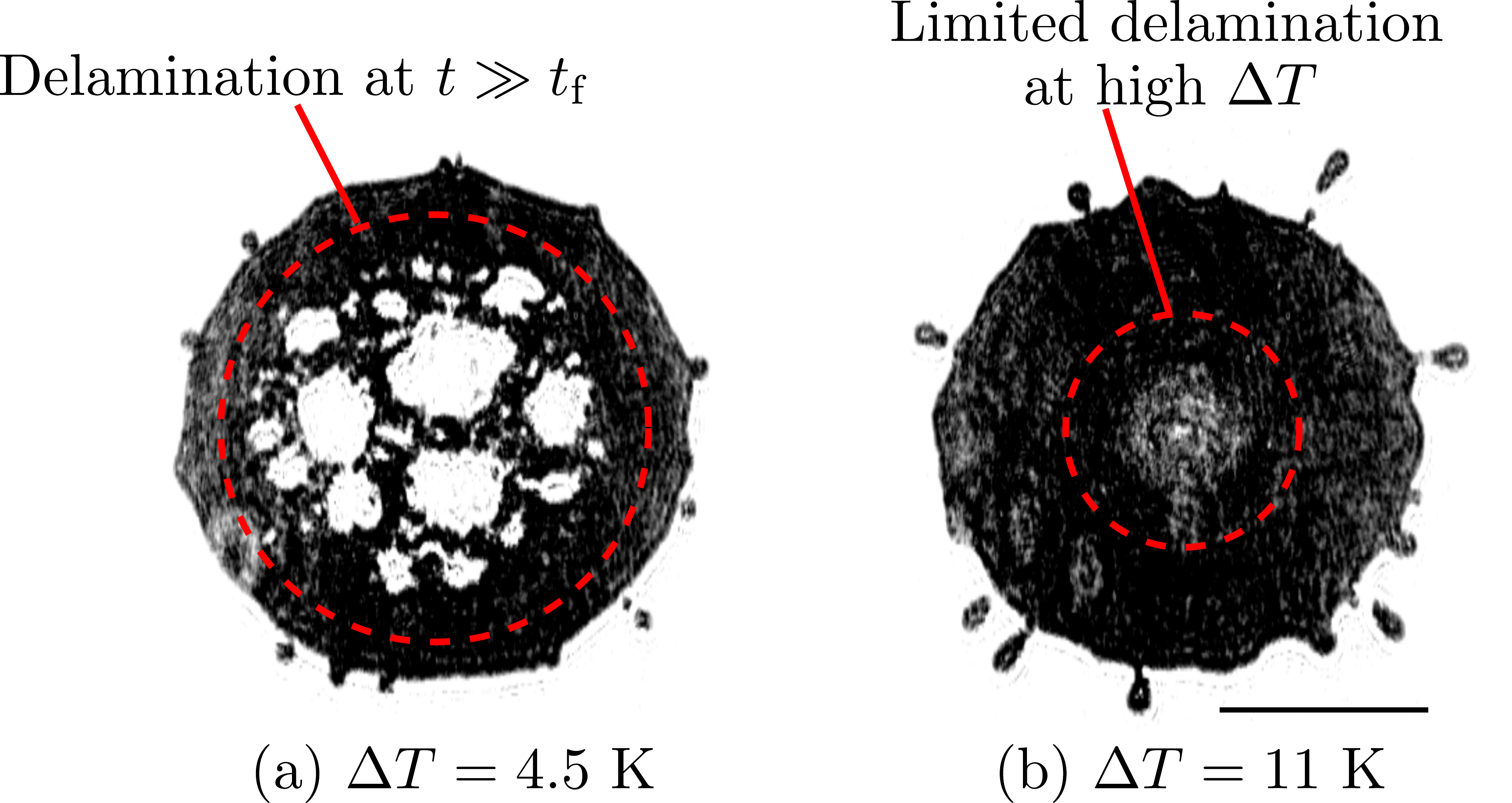}\\
\caption{Experimental snapshots showing the delamination behaviour of the frozen splat at $t \sim \mathcal{O} (10 s) > t_{\mathrm{f}}$. Contrary to the previous experimental reports for metallic droplet \cite{de2018self}, in our case the delamination predominantly occurs in the central region of the frozen splat, and the extent of delamination reduces for higher $\Delta T$. The horizontal scale bar corresponds to 2 mm. Corresponding movies (S6 and S7) are available online as supplementary material.}
\label{fig:fig4}
\end{figure}

Finally, we shift our focus to the long term behaviour of the frozen splat.
After the impact of a droplet on the undercooled substrate, it spreads out and slowly solidifies to form a thin crust of thickness $h \sim 150\,\mu$m in time $t_{\mathrm{f}}$.
At times much larger than $t_{\mathrm{f}}$, we noted that this thin solidified crust detaches from the substrate.
This behaviour is similar to the delamination of the frozen metallic splats recently reported in Ref. \cite{de2018self}.
It was reported that the delamination of tin splats initiates at the edges and creeps towards to the centre.
Moreover, the extent increases for high $\Delta T$.
The authors argued that delamination is triggered by combination of thermo-mechanical stresses and interfacial defects at the bottom of the frozen splat.
On the contrary, we noted that, at low $\Delta T$, the delamination of a hexadecane splat initiates at random locations (away from the boundaries) and eventually most of the central area of the frozen splat detaches from the substrate; see Fig. \ref{fig:fig4}.
Moreover, with increasing $\Delta T$, the extent of delamination reduces and remains limited to the central region of the frozen splat.
Since no interfacial defects were observed for the hexadecane splats, we believe that in our system the delamination of the frozen splat could be driven by a peculiar crystallisation behaviour of hexadecane.
It is known that \textsl{n}-even alkanes crystallise to a stable Triclinic crystal state through a so-called metastable Rotator phase \cite{sirota1999transient, sirota1998supercooling}.
This metastable state is a weakly ordered crystal phase and at low supercooling $\Delta T \leq 7$ K it can persist for long times \cite{sirota1999transient}.
Therefore, we believe that at low $\Delta T$, the build-up of compressive stresses due to thermal contraction is sufficient to cause interlayer movements of the Rotator crystals which eventually leads to the delamination of the splat.
In contrast, at high $\Delta T$, the transition between Rotator and highly ordered Triclinic phase takes place rapidly, thus the extent of delamination remains limited near the location of droplet impact where the entrapped bubble in the frozen splat acts as a localised defect.

\section*{Conclusion}

In summary, by adapting the total-internal-reflection technique to the freezing problem, we revealed a peculiar freezing morphology that originates from the complex interplay between the droplet scale hydrodynamics and phase-transition effects at sufficiently high substrate undercooling.
The kinetics of the advection of sequential frozen fronts observed at high $\Delta T$  is explained by combining the elements of classical nucleation theory and droplet-scale hydrodynamics.
In addition, we reported a new type of delamination behaviour of frozen splats at times $t > t_{\mathrm{f}}$, which is drastically different from typical defect triggered detachment processes that have been studied in detail \cite{yu2003delamination}.
In our system the delamination of a frozen splat is driven by the existence of a weakly ordered transient crystalline phase which allows relative movement of the solidified material under thermo-mechanical stresses.
Since the TIR technique allows direct visualisation of the nucleation events and crystal growth occurring next to the cold surface, it also can further be employed in more complex situations, such as solidification of an impacting droplet on liquid infused or patterned surfaces which potentially can potentially influence the many industrial processes.

\matmethods{

\begin{figure}
\centering
\includegraphics[clip, trim=0cm 0cm 0cm 0cm, width=0.5\textwidth]{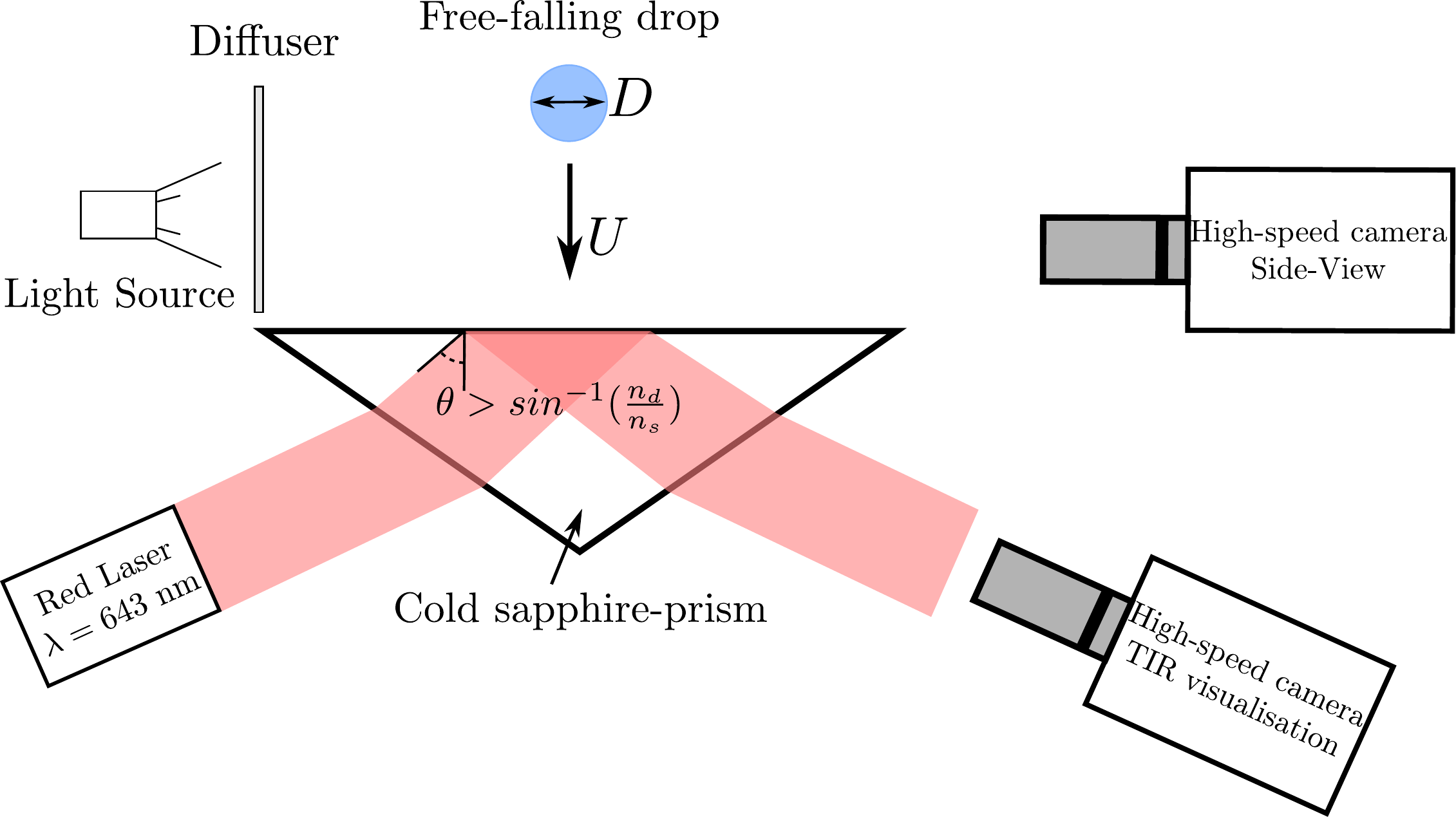}\\
\caption{Schematic diagram of the experimental setup employing total-internal-reflection technique to visualise the freezing behaviour of an impacting droplet on undercooled surface.}
\label{fig:fig5}
\end{figure}

A schematic diagram of the experimental setup is shown in Fig. \ref{fig:fig5}.
In each experiment, a pendant drop was released from the tip of a needle with an outer(inner) diameter of 240(100) $\mu$m.
The balance between surface-tension and gravity forces ensured that all droplets were of a similar size (1.58 mm $\pm$ 20 $\mu$m).
The employed liquid in the reported experiments was Hexadecane (99\%, Sigma-Aldrich), however, a few experiments were also performed with 1-Octadecene (Sigma-Aldrich) to confirm the non-exclusiveness of different freezing morphologies.
Hexadecane is optically transparent at room temperature and has a melting point of 18$^\circ$C.
The liquid has a density $\rho = 770\,\mathrm{kg/m}^3$, surface tension $\sigma = 27 \,\mathrm{mN/m}$, dynamic viscosity $\mu = 3.47 \times 10^{-3}\,\mathrm{Pa~s}$, specific heat $c_{\mathrm{p}} = 2310\,\mathrm{J/kg\,K} $, latent heat of fusion $L = 2.3 \times 10^{5}\,\mathrm{J/kg} $ and thermal diffusivity $\alpha = 8.40 \times 10^{-8}\, \mathrm{m}^2{\mathrm{/s}}$.
Here, we must emphasize that hexadecane is a non-volatile liquid and has extremely low vapour pressure $P_\mathrm{vap} = 0.308$ Pa, thus any evaporative pre-cooling effects can be ignored in our experiments.
The impact velocity $U$ was varied by adjusting the vertical distance between the horizontal substrate and the needle.
The impact event was recorded in side-view as backlit shadow-graphs and in bottom-view \textsl{via} the total-internal-reflection technique.
The impact velocity and shape/size of the droplet at the time of impact were measured from side-view images recorded with a high-speed camera (Photron APX-RS) at 10000-14000 fps with a macro lens. 
Bottom view observations (\textsl{via} TIR) were recorded using a high-speed camera (Photron Fastcam SA-X2) connected to a  long-distance microscope (Navitar 12x Telecentric zoom system) at 80000 fps. 

The horizontal surface of a sapphire prism (Crystan Ltd.) of thermal conductivity $k_\mathrm{s} = 34.60\, \mathrm{W m}^{-1}\mathrm{K}^{-1}$ was used as a model surface in the experiments.
It was placed in direct contact with a liquid-cooled aluminium holder.
The average roughness of a similar prism was measured previously \cite{shirota2016dynamic} to be less than 10 nm.
The temperature of the surface was measured before each experiment using a K-type thermocouple.
Note, that additional experiments were also performed under strongly reduced ambient pressure, ambient pressure $P_\mathrm{amb} = 25$ mbar, to understand the role of the air bubble entrapped during droplet impact and the formation of nano-bubbles during the freezing of the droplet \cite{chu2019bubble}. However, we did not measure any change in the overall kinetics of solidification due to reduced ambient pressure.

For TIR imaging, a 60 mW diode laser beam ($\lambda$ = 643 nm), expanded to $\approx 2$ cm diameter, was introduced to the prism via mirrors at a certain incident angle. 
The incident angle was carefully chosen such that it is greater than the critical angle $\theta_{\mathrm{TIR}} > \mathrm{sin}^{-1}(\frac {n_{\mathrm{d}}}{n_{\mathrm{s}}}) $, where $n_{\mathrm{d}} = 1.43$ and $n_{\mathrm{s}} = 1.76$ are the refractive indices of liquid hexadecane and sapphire, respectively. 
This ensured that in the bottom views recorded \textsl{via} TIR, the liquid in contact with the sapphire surface is not visible.
However, an evanescent wave emerges in the droplet, whose intensity decays in an exponential manner within one wavelength distance from the substrate.
Hence, when a growing crystal in the vicinity of the substrate interacts with the evanescent wave it is visible in the images.
For further details of the experimental method we refer to the supplementary material.
 
Data Availability Statement: All data discussed in the paper is made available to readers.
}

\showmatmethods{} 

\acknow{We acknowledge the funding by the Max Planck Center Twente, P. Kant acknowledges funding from NWO,  R. B. J. Koldeweij acknowledges funding by the TNO ERP program 3D nano-manufacturing, K. Harth acknowledges funding by German Science Foundation (DFG) within grant HA8467/1-1, D. Lohse acknowledges the funding from ERC Adv. grant DDD 740479.}

\showacknow{} 

\bibliography{bibliography}

\end{document}